\documentclass[12pt,a4paper]{article}

\usepackage{amsfonts}

\newcommand{\be}{\begin{eqnarray}}
\newcommand{\ee}{\end{eqnarray}}

\begin{document}

\title {On the maximum entropy principle in non-extensive thermostatistics}

\author{Jan Naudts\\
\small  Departement Fysica, Universiteit Antwerpen,\\
\small  Universiteitsplein 1, 2610 Antwerpen, Belgium\\
\small E-mail: Jan.Naudts@ua.ac.be.
}

\date {May 2004}

\maketitle

\begin{abstract}
It is possible to derive the maximum entropy principle from
thermodynamic stability requirements.
Using  as a starting point the equilibrium probability distribution, currently used in
non-extensive thermostatistics, it
turns out that the relevant entropy function
is R\'enyi's $\alpha$-entropy, and not Tsallis' entropy.
\end {abstract}

\section {Introduction}

The maximum-entropy principle has played an essential role in the development
of non-extensive thermostatistics, starting with Tsallis' paper \cite {TC88}
in 1988. The Tsallis formalism has known several major revisions without reaching a stable
and satisfactory status. Origin of the difficulties might be that the
maximum-entropy principle is used as a postulate of statistical mechanics,
while it is rather a result that can be derived from first principles.
Below is used that the maximum-entropy principle is a consequence of the
statement that free energy is minimal at equilibrium. The definition of
what equilibrium is, what free energy is, comes first. The maximum-entropy principle
follows. This point of view
differs from that of Tsallis literature where equilibrium is by definition the state maximizing
some entropy functional under certain constraints.

This change of viewpoint has some important consequences. Given a family of temperature
dependent probability distributions one can calculate the entropy functional which
is needed to reproduce the given probability distributions. In the present paper the
probability distributions used in recent Tsalllis literature are taken as starting
point. Surprisingly, the resulting entropy functional is {\sl not} Tsallis entropy
but is the $\alpha$-entropy of R\'enyi. This result agrees with the known fact
\cite {LMdS00} that the maximum entropy principle in combination
with R\'enyi's entropy reproduces the equilibrium probability distributions of
non-extensive thermostatistics.
Note that there exists an extensive literature on R\'enyi's $\alpha$-entropies.
Some of the references in the context of non-extensive thermostatistic
are \cite {LMdS00,NC02,BAG04,JA04}.

The next section serves to fix notations and starts with the probability distributions
that are found as equilibrium states in recent Tsallis literature.
Section 3 introduces thermodynamic entropy $S(U)$ as a function of average energy $U$.
It is obtained from average energy by integrating the well-known thermodynamic definition of temperature.
In Section 4 the requirement of thermodynamic stability is worked out.
In Section 5 the maximum entropy principle is derived. The final section contains
a short discussion of results. Some of the calculations are given in Appendix.

\section {Equilibrium probability distributions}

For sake of simplicity notations refer to a classical gas of $n$ particles
enclosed in a container of volume $V$ in $\nu$-dimensional space. The energy functional
is denoted  $H(x)$, with $x$ a point of phase space $\Gamma$.

Average energy\be
U(\beta)=\int_\Gamma{\rm d}x\, p_\beta(x)H(x)
\ee
is calculated using the distribution \cite {TMP98,MNPP00}
\be
p_\beta (x)=\frac 1{Z(\beta)}\left[1-(1-q)\beta[H(x)-U(\beta)]\right]_+^{q/(1-q)}.
\label {pdfabe}
\ee
In this expression $Z$ is the normalization
\be
Z(\beta)=\int_{\Gamma}{\rm d}x\,\left[1-(1-q)\beta[H(x)-U(\beta)]\right]_+^{q/(1-q)}.
\ee
The notation $[u]_+=\max\{0,u\}$ is used.
Equation (\ref {pdfabe}) can be written as
\be
p_\beta (x)=\frac 1{Z^*(\beta^*)}\left[1-(1-q)\beta^*H(x)]\right]_+^{q/(1-q)}
\ee
with
\be
\beta^*=\frac \beta{1+(1-q)\beta U(\beta)}.
\label {imp1}
\ee
In the latter form normalization is straightforward and average energy can be calculated
\be
U(\beta)=\frac 1{Z^*(\beta^*)}\int_{\Gamma}{\rm d}x\,
\left[1-(1-q)\beta^*H(x)]\right]_+^{q/(1-q)}H(x)
\label {imp2}
\ee
The pair of equations (\ref {imp1}, \ref {imp2}) should have a unique solution $U(\beta)$.
In addition, this solution should be a decreasing function of $\beta$. See Appendix A
for a discussion of the existence of solutions.

\section {Thermodynamic entropy}

The basic thermodynamic quantity is entropy $S(U)$ as a function of energy $U$.
It is used to define inverse temperature $\beta$ by
\be
\beta=\frac {{\rm d}S}{{\rm d}U}.
\label {temp}
\ee
In the present context $U(\beta)$ is given. Hence $S(U)$ can be obtained
by a simple integration of (\ref {temp}).
This is done in Appendix B. The result is
\be
S(U(\beta))=\ln\left[\left(\frac mh\right)^{\nu n} Z(\beta)\right].
\label {neS}
\ee
Here,
$m$ is the mass of each particle, and $h$ is a constant inserted for dimensional reasons
(units are used in which $k_B=1$).
The integration constant has been chosen so that for the ideal gas
in the limit $q=1$ one obtains the standard result
\be
S(U)=\frac {\nu n}2\ln\frac {2\pi em}{h^2}  U V^{2/\nu}
-\frac {\nu n}2\ln \frac {\nu n}2.
\label {idealS}
\ee

\section {Thermodynamic stability}

Thermodynamic stability of equilibrium states requires \cite {CHB85}
in the first place that
\be
\frac {\partial^2 S}{\partial U^2}\le 0.
\label {stab1}
\ee
This condition boils down to the requirement that $U(\beta)$ is a decreasing function of $\beta$.
This is assumed to be the case, but cannot be proved under general conditions for the
family of probability distributions (\ref {pdfabe}).

If (\ref {stab1}) is satisfied for all $U$ then $S(U)$ can be replaced by its
Legendre transform, which is called Helmholtz free energy $F$,
and which historically is defined by
\be
F=U-TS(U)
\ee
with temperature $T$ given by $T=1/\beta$. Free energy $F$ is a concave function of temperature $T$.
Its first derivative satisfies
\be
\frac {{\rm d}F}{{\rm d}T}=-S.
\ee

Thermodynamic stability is more than just (\ref {stab1}). The system should
be stable against all possible perturbations. For that reason the
free energy should be concave, not only as a function of temperature
but in a much larger parameter space. To formulate this requirement
in a consistent way it is easier to work with $\beta F=\beta U-S$
instead of with $F$. Note that
\be
\frac {{\rm d}\,}{{\rm d}\beta}\beta F=U.
\ee
Hence the condition of concavity of $F(T)$ is equivalent with concavity of $\beta F$
as a function of $\beta$. The quantity $\beta F$ depends on both $\beta$ and on the
choice of energy function $H(x)$, but only the combination $\beta H(x)$ occurs. Hence
the requirement that $\beta F$ is concave in $\beta$ can be extended to the requirement
that $\beta F$ is a concave function over the space of all possible $\beta H(x)$.
The latter requirement is too strong, because of its global nature,
and should be weakened by requiring local concavity
of $\beta F$. A tangent plane to $\beta F$ in the point $\beta H$ is given by the linear map
\be
\beta' H'\rightarrow \beta F+\beta'\langle H'\rangle-\beta\langle H\rangle
\ee
where averages are equilibrium averages w.r.t.~$\beta,H$. The requirement for stability
of the system with energy functional $H(x)$ at inverse temperature $\beta$
is therefore that
\be
\beta'F'\le \beta F+\beta'\langle H'\rangle-\beta\langle H\rangle
\label {stable}
\ee
for all possible choices of $\beta',H'$.
Note that (\ref {stable}) can be written as
\be
S'-\beta' U'\ge S-\beta \langle H'\rangle.
\label {entineq}
\ee
In this expression, the averages $\langle\cdot\rangle$ are still calculated
w.r.t.~the equilibrium probability distribution of the system with energy
functional $H(x)$ at inverse temperature $\beta$.

\section {Maximum entropy principle}

The origin of the maximum entropy principle lies in the requirement of
thermodynamic stability. Indeed, (\ref {entineq}) can be read as the statement
that the quantity $S-\beta \langle H'\rangle$ is maximal when $\beta=\beta'$
and $H=H'$, i.e., when for the calculation of free energy the correct
probability distribution is used.
This is the maximum entropy principle.
In the canonical formalism of statistical physics, based on the Boltzmann-Gibbs
distribution, systems with a finite number of degrees of freedom are generically
stable. Hence it is acceptable to interchange primed and unprimed systems.
In the more general context of non-extensive thermostatistics this interchange
is not acceptable, since it is known (see e.g.~\cite {AS99}) that instabilities can occur.
One concludes that the maximum entropy principle holds only under the condition
that the variation of $S-\beta \langle H'\rangle$ is restricted to those pairs
of inverse temperature $\beta$
and energy functional $H(x)$ which are thermodynamically stable.
Of course, this statement is of an academic rather than practical value.
But it shows that maximum entropy should not be introduced in an axiomatic
manner. It also gives somewhat more freedom in deriving a variational
principle which is useful for practical purposes because it is not
necessary to prove that the variational solution corresponds with
a maximum. Indeed, in principle the variation should be
restricted to the equilibrium distributions of stable systems, and
by definition of stability these will have a lower free energy than the
variational solution.

Let us now introduce an entropy functional which reproduces (\ref {pdfabe})
by means of the maximum entropy principle.
Inspection of (\ref {neS}) learns that the correct {\sl ansatz}
for entropy as a function of probability distribution is
\be
I(p)=-\ln \left(\frac mh\right)^{\nu n}
\left[\int_{\Gamma}{\rm d}x\,p(x)^{1/q}\right]^{q/(1-q)}
\ee
This expression is proportional to R\'enyi's $\alpha$-entropy with $\alpha=1/q$.
One verifies immediately, using the expressions from Section 2,
that $I(p_{\beta})=S(U(\beta))$.
Variation of
\be
{\cal L}(p)=I(p)-\beta\int_{\Gamma}{\rm d}x\,p(x)[H(x)-U(\beta)]
-\alpha\left[\int_{\Gamma}{\rm d}x\,p(x)-1\right]
\ee
gives
\be
p(x)=\frac 1{Z(\beta)}
\left[-(1-q)\alpha-(1-q)\beta[H(x)-U(\beta)]\right]_+^{q/(1-q)}
\label {pdfrenyi}
\ee
with
\be
Z(\beta)=\left[\int_{\Gamma}{\rm d}x\,p(x)^{1/q}\right]^{-q/(1-q)}
\ee
Integrating $p(x)^{1/q}$, as given by (\ref {pdfrenyi}), gives
\be
Z(\beta)
&=&\int_{\Gamma}{\rm d}x\,
\left[-(1-q)\alpha-(1-q)\beta[H(x)-U(\beta)]\right]_+^{1/(1-q)}\cr
&=&-(1-q)\alpha\int_{\Gamma}{\rm d}x\,
\left[-(1-q)\alpha-(1-q)\beta[H(x)-U(\beta)]\right]_+^{q/(1-q)}.\cr
& &
\ee
Hence $\alpha=-1/(1-q)$ results in a correctly normalized
probability distribution. This shows that (\ref {pdfrenyi}) coincides with (\ref {pdfabe}).
One concludes that there exists a maximum entropy principle which produces
(\ref {pdfabe}) as the equilibrium probability distributions and
is such that the attained maximum value coincides with thermodynamic entropy.

\section {Discussion}

This paper contains one message and one result. The message is that the maximum entropy principle
is a consequence of the requirement of thermodynamic stability. In the context of
the Boltzmann-Gibbs distribution this is an evidence. In that case systems
with a finite number of degrees of freedom are automatically stable. This allowed
Jaynes \cite {JET89} to attribute an axiomatic status to the maximum entropy principle.
In the current version of Tsallis thermostatistics systems with a finite number of degrees
of freedom are {\sl not} automatically stable. The conclusion of the present paper
is therefore that maximum entropy should not be used in an axiomatic manner,
but should be derived from the requirement of thermodynamic stability.

The result of the present paper is that the equilibrium probability distributions,
as found in recent Tsallis literature, can be derived by means of the
maximum entropy principle. However, the relevant entropy functional
is R\'enyi's $\alpha$-entropy rather than Tsallis' entropy functional.
Of course, there is an easy connection between the two entropies \cite {CT91}.
But the present paper shows that a thermodynamically correct definition of temperature
leaves no choice but to use R\'enyi's entropy instead of that of Tsallis.

Mainstream non-extensive thermostatistics is based on a variational principle
which involves escort probability distributions \cite {TMP98,MNPP00}. The main
objection, from the present point of view, against these papers is that they do not reproduce
thermodynamic entropy (\ref {neS}). As a
consequence, the mainstream derivation of (\ref {pdfabe}) has problems with a correct definition
of temperature. Bashkirov, in the conclusions of his paper \cite {BAG04}, hopes that one
"can abandon the use of escort probabilities for calculations
of average values in favor of original distributions."
This statement seems to be exaggerated. We know that escort probabilities are important in
fractal physics \cite {BS93}. As shown in \cite {NJ03,NJ04,NJ04b} they also
play an important role in generalized thermostatistics.
It only happens that these escort probabilities are not involved in
the formulation of the maximum entropy principle.

\section*{Appendix A}

 \renewcommand{\theequation}{A\arabic{equation}}
  \setcounter{equation}{0}  

In this appendix the solution of (\ref {imp1}, \ref {imp2}) is discussed.
Elimination of $U(\beta)$ gives
\be
\frac 1{\beta^*}&=&\frac 1\beta +(1-q)U(\beta)\cr
&=&\frac 1\beta +\frac 1{\beta^*}\frac 1{Z^*(\beta)}\int_{\Gamma}{\rm d}x\,
\left[1-(1-q)\beta^*H(x)]\right]_+^{q/(1-q)}(1-q)\beta^*H(x)\cr
&=&\frac 1\beta+\frac 1{\beta^*}
-\frac 1{\beta^*}\frac 1{Z^*(\beta)}\int_{\Gamma}{\rm d}x\,
\left[1-(1-q)\beta^*H(x)]\right]_+^{1/(1-q)}.
\ee
This simplifies to
\be
\beta^*=\beta\frac
{\int_{\Gamma}{\rm d}x\,\left[1-(1-q)\beta^*H(x)]\right]_+^{1/(1-q)}}
{\int_{\Gamma}{\rm d}x\,\left[1-(1-q)\beta^*H(x)]\right]_+^{q/(1-q)}}.
\ee
See \cite {NJ00} for methods to prove the existence and uniqueness of the
solution $\beta^*$ of this equation.
Then average energy is given by
\be
U(\beta)=\frac 1{1-q}\left(\frac 1{\beta^*}-\frac 1\beta\right).
\ee
Thermodynamic stability requires that $U(\beta)$ is a decreasing function of
$\beta$. There is no general argument why this should always be the case.

\section*{Appendix B}

 \renewcommand{\theequation}{B\arabic{equation}}
  \setcounter{equation}{0}  

In this appendix entropy $S(U)$ is calculated.
One has
\be
0&=&(1-q)\beta\left[\langle H\rangle-U(\beta)\right]\cr
&=&\frac 1{Z(\beta)}\int_{\Gamma}{\rm d}x\,
\left[1-(1-q)\beta[H(x)-U(\beta)]\right]_+^{q/(1-q)}
(1-q)\beta[H(x)-U(\beta)]\cr
&=&-\frac 1{Z(\beta)}\int_{\Gamma}{\rm d}x\,
\left[1-(1-q)\beta[H(x)-U(\beta)]\right]_+^{1/(1-q)}
+1.
\ee
Hence one obtains
\be
Z(\beta)=\int_{\Gamma}{\rm d}x\,\left[1-(1-q)\beta[H(x)-U(\beta)]\right]_+^{1/(1-q)}.
\label {Zrel}
\ee
From the latter expression follows
\be
\frac{{\rm d}\,}{{\rm d}\beta}\ln Z(\beta)
&=&-\frac 1{Z(\beta)}\int_{\Gamma}{\rm d}x\,\left[1-(1-q)\beta[H(x)-U(\beta)]\right]_+^{q/(1-q)}\cr
& &\times
\left[H(x)-U(\beta)-\beta\frac {{\rm d}\,}{{\rm d}\beta}U(\beta)\right]\cr
&=&\beta\frac {{\rm d}\,}{{\rm d}\beta}U(\beta)\cr
&=&\frac {{\rm d}\,}{{\rm d}\beta}S(U(\beta)).
\ee
In the last line (\ref {temp}) has been used.
Integration gives
\be
S(U(\beta))=\ln Z(\beta)+\hbox{ constant}.
\ee
This is  (\ref {neS}), up to the integration constant.

Let us now show that (\ref {neS}) in case of the ideal gas and in the limit $q=1$
reduces to (\ref {idealS}).
For the ideal gas, in the limit $q=1$, is
\be
Z(\beta)
&=&e^{\beta U(\beta)}\int_{\Gamma}{\rm d}x\,e^{-\beta H(x)}\cr
&=&e^{\beta U(\beta)}V^n\left(\frac {2\pi}{\beta m}\right)^{\nu n/2}.
\ee
Therefore, (\ref {neS}) becomes
\be
S(U(\beta))&=&\ln\left[\left(\frac mh\right)^{\nu n}
e^{\beta U(\beta)}V^n\left(\frac {2\pi}{\beta m}\right)^{\nu n/2}
\right].
\ee
This yields (\ref {idealS}) because $\beta U(\beta)=\nu n/2$.


\end{document}